\documentclass[runningheads]{llncs}
\usepackage[T1]{fontenc}
\usepackage{graphicx}
\usepackage{listings}
\usepackage{xcolor}
\usepackage{bbding}
\usepackage{amsmath, amssymb, bm}
\usepackage[boxed,vlined]{algorithm2e}

\usepackage{booktabs}

\begin{document}
\title{Spatio-Temporal Scheduling Prediction Under Backhaul Delay for Resilient Coordinated Beamforming}
\titlerunning{Resilient Coordinated Beamforming Under Backhaul Delay}

\author{Prashant Kumar Singh\Envelope\inst{1,2}\orcidID{0009-0006-4222-678X} \and
Shubham Vaishnav\inst{1}\orcidID{0000-0001-7612-4227} \and
Ahmet Hasim G\"{o}kceoglu\inst{2}\orcidID{0000-0002-2706-1079} \and
Li Wang\inst{2}\orcidID{0000-0003-3365-2403}}
\authorrunning{P. K. Singh et al.}
\institute{Stockholm University, Stockholm, Sweden\\
\email{\{prku7110, shubham.vaishnav\}@dsv.su.se}
\and
Huawei R\&D, Stockholm, Sweden\\
\email{\{ahmet.hasim.gokceoglu1, leo.li.wang\}@huawei.com}}
\maketitle

\begin{abstract}
Coordinated beamforming in distributed 5G networks relies on the timely
exchange of inter-cell scheduling information, but backhaul latency
makes this information stale. Even a single transmission time interval
(TTI) of delay can reduce CBF-SLNR performance below the uncoordinated
baseline, because the precoder suppresses interference toward users
that are no longer active. Coordination on stale information is
therefore worse than no coordination at all. To address this, we
propose a two-stage predictive framework in which a Spectral Temporal
Graph Neural Network (StemGNN) predicts future user equipment (UE)
scheduling states from delayed historical observations, and the
predictions replace stale inputs to the CBF-SLNR precoder. Evaluated
on a three-cell massive MIMO downlink with 60 UEs and 64 antennas per
base station under Quadriga Urban Micro (UMi) channels and a
proportional fair scheduler, StemGNN achieves a mean scheduling
prediction accuracy of 87.57\%, outperforming LSTM, GRU, Simple RNN,
and Markov chain baselines at all evaluated horizons, with gains of up
to 7.71\% over LSTM at longer horizons where inter-UE structural
dependencies dominate over temporal autocorrelation. When integrated
into coordinated beamforming, the predictions recover 57--73\% of the
sum rate loss caused by one TTI of backhaul delay, improving sum rate
by 9.58--14.35\% over the no-prediction baseline and recovering up to
83\% of the Lag-1 fairness loss for cell-edge users, with fairness
gains persisting at higher lag values where throughput gains diminish.
These results show that treating backhaul latency as a spatio-temporal
forecasting problem is an effective approach for robust inter-cell
coordination in delay-constrained networks.

\keywords{Coordinated beamforming \and StemGNN \and Scheduling prediction \and Backhaul delay \and Edge AI}
\end{abstract}

\section{Introduction}
\label{sec:introduction}

Modern 5G networks, and emerging 6G ones, use coordinated beamforming (CBF) across multiple base stations (BSs) to reduce inter-cell interference and keep spectral efficiency high~\cite{Gesbert,wmmse,dahlman,Larsson}. In distributed radio access networks, neighboring BSs exchange channel state and scheduling information over backhaul links. These links have limited bandwidth, are often asynchronous, and frequently add delay~\cite{park2013dynamic,checko2015cloudran}. As a result, the view each BS has of the wider network is almost always out of date, while user activity changes much faster: users enter, leave, or change their scheduling state continuously~\cite{tse2005wireless,kelly1997proportionalfair}. Standard CBF algorithms such as Weighted Minimum Mean Square Error (WMMSE) optimization~\cite{wmmse,fp} assume that each BS has fresh, global channel and scheduling information. Once this assumption breaks, the sum rate drops and service quality becomes uneven across cells.

From an edge AI point of view, each BS is an autonomous edge node that has to make real-time, latency-critical decisions using incomplete and partially outdated information about its peers. Backhaul latency, jitter, and asynchronous updates act as a kind of partial fault on the links between nodes: the information a BS needs is not lost outright, it just arrives late and stale. A beamforming controller at the edge therefore needs to stay robust against this kind of information-staleness fault to keep service available and quality of experience consistent across cells.

The trust side of this problem matters too. Each BS feeds its beamformer with scheduling and channel reports it receives from neighbors, and there is no built-in way to tell whether an unusual or late-arriving report reflects genuine network behavior, a backhaul fault, or a tampered message injected somewhere along the path. The more directly a BS depends on raw, last-received neighbor data, the larger this trust surface becomes. A learning-based predictor trained on the legitimate temporal patterns of neighbor scheduling reduces this dependence: the BS can rely on its own predicted view of the neighborhood, and treat late-arriving reports as a correction rather than as ground truth. We do not aim to detect adversarial inputs in this work, but the same prediction step that handles delay also shrinks the attack surface exposed to an attacker on the backhaul.

Despite a lot of progress on CBF and on machine-learning-assisted interference management, most existing work either assumes ideal information exchange, or treats the spatial and temporal sides of the problem separately. Graph Neural Networks (GNNs) capture the spatial structure of wireless networks well by representing BSs and users as nodes in a graph~\cite{zhou2020gnnreview,kipf2017gcn,gilmer2017mpnn}, and recent GNN-based beamforming methods~\cite{he,8938771,miso,10256678} perform strongly but usually assume static user configurations and synchronous information access. On the other side, spectral-temporal forecasting models such as StemGNN~\cite{cao,wu2019graphwavenet} capture multivariate temporal dependencies but are not connected to physical-layer beamforming decisions. As a result, no existing framework jointly addresses (i) delayed inter-BS information, (ii) a dynamic user population, and (iii) coordinated beamforming in a form that fits an edge deployment.

\begin{figure}[t]
  \centering
  \includegraphics[width=\textwidth]{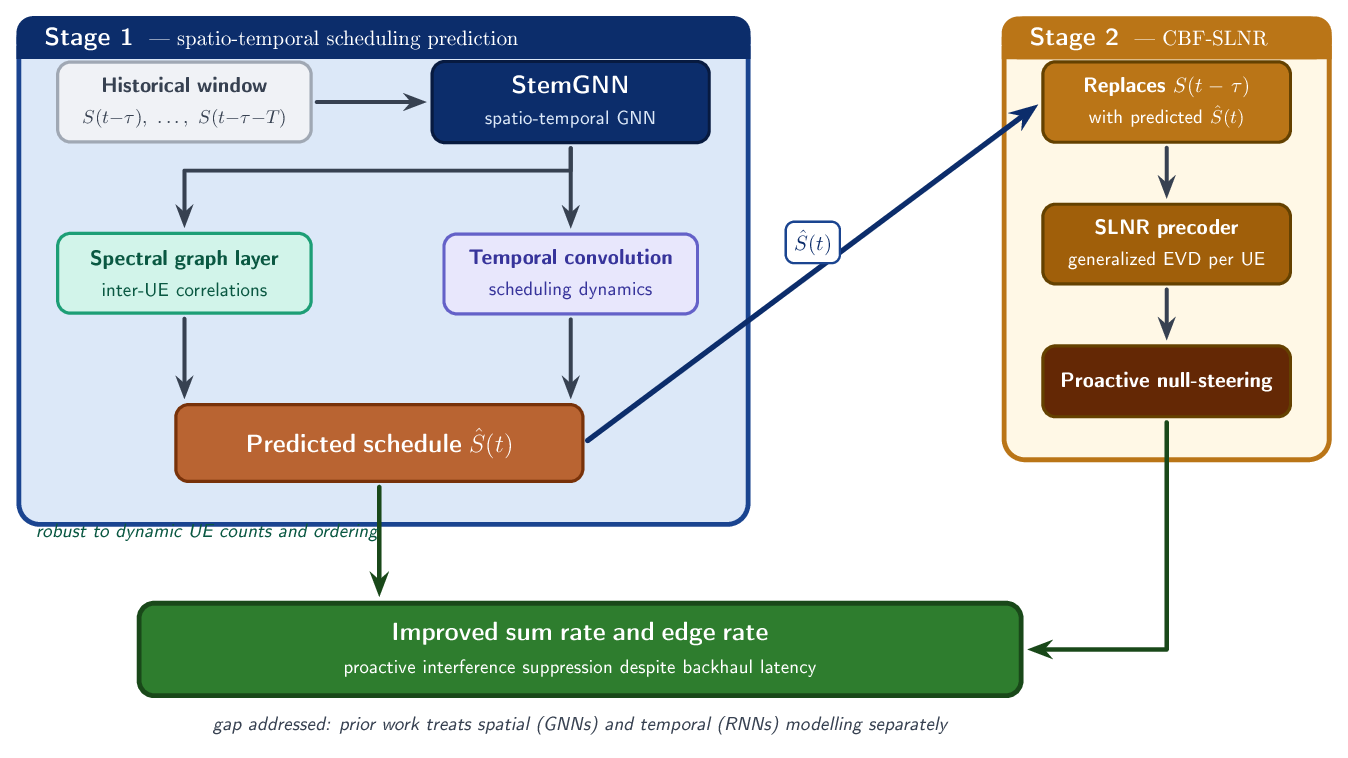}
  \caption{Proposed prediction-assisted coordinated beamforming framework. Each base station uses a spectral-temporal GNN to predict the current scheduling state of neighboring cells from delayed backhaul observations, and feeds the predictions into an SLNR-based coordinated beamformer.}
  \label{fig:solution}
\end{figure}

In this paper, we recast coordinated beamforming under backhaul delay as a prediction-assisted decision problem that runs locally at each BS. Each edge node uses a spectral-temporal GNN, based on StemGNN~\cite{cao}, to predict the current user-equipment (UE) scheduling state of neighboring cells from past, delayed observations, and feeds these predictions into a Signal-to-Leakage-plus-Noise Ratio (SLNR) coordinated beamformer (see Fig.~\ref{fig:solution} for an overview). To handle the dynamic user populations seen in real networks, we extend the architecture to support a variable number of UEs and a permutation-invariant representation, so the same model still works as users join, leave, or are reordered. The resulting controller suppresses inter-cell interference even when the backhaul is delayed and user activity is stochastic, and gives a concrete example of a fault-tolerant edge AI pipeline for cooperative wireless inference.

\noindent\textbf{Contributions.} The main contributions of this paper are as follows:
\begin{itemize}
  \item We formulate inter-BS UE scheduling prediction under backhaul delay as a multivariate binary time-series problem, and benchmark a spectral-temporal GNN against recurrent (RNN, LSTM, GRU) and Markov baselines.
  \item We extend StemGNN with a permutation-invariant input and output representation that supports a dynamic, variable number of UEs without retraining, making it suitable for realistic edge deployments.
  \item We integrate the predicted scheduling states into an SLNR-based CBF pipeline and measure the resulting sum-rate gain over conventional beamforming that uses stale information directly, under a range of backhaul delays and dynamic user-activity conditions.
\end{itemize}

\section{Related Work}
\label{sec:related_work}

\subsection{Coordinated beamforming and the cost of stale information}

Massive MIMO and coordinated beamforming have become the standard tools for managing interference in dense multi-cell deployments~\cite{Marzetta,Larsson,Lu,Boccardi,Gesbert}. The classical algorithms in this area, mainly WMMSE~\cite{wmmse} and fractional programming~\cite{fp}, reach near-optimal sum rate when each base station has accurate and timely channel state and scheduling information from all of its neighbors. In practice this is rarely the case. In cloud and distributed RAN architectures, backhaul links are bandwidth-limited and asynchronous, and the inter-cell information used to drive these algorithms is delayed, partial, or both~\cite{checko2015cloudran,park2013dynamic}. The repeated matrix inversions and iterative steps in WMMSE and FP also push their computational cost beyond what is reasonable for real-time execution on an edge node such as a base station~\cite{Lu,Marzetta}. Decentralized variants reduce some of this burden, but still require frequent inter-cell signaling, which is affected by the same backhaul delays~\cite{Gesbert,li}. From an edge-deployment standpoint, this points to a clear failure mode: when the data feeding the coordination algorithm is late or incomplete, the algorithm produces a confidently wrong decision, and the system has no fallback.

\subsection{Machine learning approaches and their edge implications}

To work around these limitations, a large body of recent work uses machine learning for wireless resource allocation and interference management~\cite{Farsad2018deepwireless,Zappone2019deepopt,Giordani2020towards6gAI}. The general idea is to replace iterative optimization with a learned mapping from network state to control actions, which lowers inference latency and fits the timing budget of edge execution. Reinforcement learning has been applied to beamforming, power allocation, and inter-cell coordination by framing them as sequential decision problems~\cite{Luong2019deepRLsurvey,Nguyen2020multiagentwireless}, though stability under fast-changing conditions remains a concern. Supervised deep-learning models that map channel observations directly to beamforming vectors~\cite{Sun2018learningoptimization} cut complexity further, but tend to ignore the structural dependencies between cells. From a deployment standpoint, these models also raise questions about trust and robustness: a base station that runs a learned model at the edge inherits the integrity of its training data and the inputs it consumes at runtime, and most of these works do not discuss how the model behaves when those inputs are stale, missing, or tampered with.

\subsection{Graph and spatio-temporal models}

Wireless networks have a natural graph structure, with base stations and users as nodes and interference relationships as edges. Graph Neural Networks (GNNs)~\cite{kipf2017gcn,gilmer2017mpnn,Wu2020gnnsurvey,zhou2020gnnreview} are a good fit for this structure because their message passing is permutation invariant and respects the connectivity of the network. GNN-based beamforming methods, including heterogeneous variants such as HSTGNN~\cite{he,Zhang2022gnnwireless} and the approaches in~\cite{8938771,miso,10256678}, achieve near-optimal sum rate with millisecond-level inference, which is compatible with edge execution. The main limitation across this line of work is the assumption of static, or slowly varying, user configurations. Real cellular traffic does not behave this way: users enter, leave, and change scheduling state on much shorter timescales~\cite{dahlman,tse2005wireless,kelly1997proportionalfair}.

On the temporal side, spectral-temporal models such as StemGNN~\cite{cao} forecast multivariate time series by jointly modeling temporal dependencies and inter-series correlations in the frequency domain. Graph WaveNet~\cite{wu2019graphwavenet} combines graph convolution with dilated temporal convolutions for a similar purpose, and DCRNN and STGCN follow related ideas~\cite{Li2018dcrnn,yan2018spatial}. These models are designed as forecasters, and are not typically connected to a physical-layer decision such as beamforming.

\subsection{Positioning of this paper}

Pulling these threads together, three observations stand out for an edge AI deployment in a distributed RAN. Classical CBF assumes information conditions that the edge does not actually have. ML-based beamforming captures spatial structure but treats user activity as static. Spatio-temporal forecasters capture the temporal side but stop short of the beamforming pipeline. To the best of our knowledge, no existing framework jointly addresses (i) the spatial structure of the network, (ii) the temporal evolution of user activity, and (iii) the staleness of inter-base-station information, in a form deployable at the edge. This paper targets exactly that intersection by combining a spectral-temporal GNN predictor for delayed neighbor scheduling with an SLNR-based coordinated beamformer, so that each base station can take a beamforming decision without trusting a possibly faulty or stale neighbor report at face value.


\section{System Model and Problem Formulation}
\label{sec:formulation}

\subsection{Multi-Cell Downlink Model}

We consider a downlink Massive MIMO network with $B$ base stations (BSs) operating on the same time--frequency resource, where each BS $b \in \{1, \dots, B\}$ has $M$ antennas and serves $K_b$ single-antenna user equipments (UEs). The total number of UEs in the network is $K = \sum_{b=1}^{B} K_b$. We denote the channel vector from BS $b'$ to UE $k$ served by BS $b$ as $\mathbf{h}_{b',b,k} \in \mathbb{C}^{M}$. In our experiments, these channels are generated with the Quadriga Urban Micro propagation model~\cite{dahlman}, which captures path loss, shadowing, and multipath fading.

At time slot (TTI) $t$, BS $b$ transmits the linearly precoded signal
\begin{equation}
\mathbf{x}_b(t) = \sum_{k=1}^{K_b} \mathbf{w}_{b,k}(t)\, s_{b,k}(t),
\label{eq:tx_signal}
\end{equation}
where $\mathbf{w}_{b,k}(t) \in \mathbb{C}^{M}$ is the beamforming vector for UE $k$ in cell $b$ and $s_{b,k}(t) \sim \mathcal{CN}(0,1)$ is the unit-power data symbol. Each BS is subject to a per-cell transmit power constraint $\sum_{k=1}^{K_b} \|\mathbf{w}_{b,k}(t)\|^2 \le P_{\max}$.

The signal received at UE $k$ in cell $b$ is
\begin{equation}
y_{b,k}(t) = \mathbf{h}_{b,b,k}^{H} \mathbf{w}_{b,k} s_{b,k}
 + \sum_{j \neq k} \mathbf{h}_{b,b,k}^{H} \mathbf{w}_{b,j} s_{b,j}
 + \sum_{l \neq b}\sum_{j=1}^{K_l} \mathbf{h}_{l,b,k}^{H} \mathbf{w}_{l,j} s_{l,j}
 + u_{b,k},
\label{eq:rx_signal}
\end{equation}
where $u_{b,k} \sim \mathcal{CN}(0, \sigma_u^2)$ is additive Gaussian noise (time index suppressed for readability). The three sums on the right-hand side are, respectively, the desired signal, the intra-cell interference, and the inter-cell interference. The resulting SINR and achievable rate at UE $(b,k)$ are
\begin{equation}
\gamma_{b,k}(t) =
\frac{|\mathbf{h}_{b,b,k}^{H} \mathbf{w}_{b,k}|^2}
{\sum_{j \neq k} |\mathbf{h}_{b,b,k}^{H} \mathbf{w}_{b,j}|^2
+ \sum_{l \neq b}\sum_{j} |\mathbf{h}_{l,b,k}^{H} \mathbf{w}_{l,j}|^2
+ \sigma_u^2},
\quad
R_{b,k}(t) = \log_2\big(1 + \gamma_{b,k}(t)\big).
\label{eq:sinr_rate}
\end{equation}

\subsection{User Scheduling and Backhaul Delay}

In each TTI, BS $b$ runs a proportional-fair scheduler~\cite{kelly1997proportionalfair,tse2005wireless} that selects a subset of its UEs for transmission. We represent this decision by the binary scheduling vector
\begin{equation}
\mathbf{s}_b(t) \in \{0,1\}^{K_b},
\quad
s_{b,k}(t) = 1 \iff \text{UE } (b,k) \text{ is active at TTI } t.
\label{eq:sched}
\end{equation}
Only active UEs contribute to the interference terms in (\ref{eq:rx_signal}), so the global scheduling state $\mathbf{s}(t) = [\mathbf{s}_1(t); \dots; \mathbf{s}_B(t)]$ directly shapes the interference pattern at every UE in the network.

In a distributed RAN, each BS shares its scheduling decisions and channel state with its neighbors over a backhaul link. We model the backhaul as introducing a fixed delay of $\tau \ge 1$ TTIs. At TTI $t$, each BS therefore observes only
\begin{equation}
\mathbf{s}_{b'}(t-\tau), \quad b' \neq b,
\label{eq:stale_obs}
\end{equation}
about its neighbors. From an edge-deployment perspective, this delay is a structural property of the link between edge nodes, not an occasional event, so any coordination logic running at the BS has to tolerate it by design.

\subsection{Coordinated Beamforming with Stale Information}

Coordinated beamforming aims to maximize the network sum rate under per-BS power constraints:
\begin{equation}
\max_{\{\mathbf{w}_{b,k}\}} \;\; \sum_{b=1}^{B} \sum_{k=1}^{K_b} R_{b,k}(t)
\quad \text{s.t.} \quad \sum_{k=1}^{K_b} \|\mathbf{w}_{b,k}\|^2 \le P_{\max}, \;\; \forall b.
\label{eq:cbf_obj}
\end{equation}
This problem is non-convex; iterative methods such as WMMSE~\cite{wmmse} and fractional programming~\cite{fp} solve it under the assumption that each BS has timely access to the global channel and scheduling state. Coordinated beamforming via the Signal-to-Leakage-plus-Noise Ratio (SLNR) criterion offers a closed-form per-UE alternative that is well suited to distributed execution at the edge~\cite{Gesbert}. For UE $k$ in cell $b$, the SLNR is
\begin{equation}
\mathrm{SLNR}_{b,k}(t) = \frac{|\mathbf{h}_{b,b,k}^{H} \mathbf{w}_{b,k}|^2}
{\mathbf{w}_{b,k}^{H} \mathbf{Q}_b(t)\, \mathbf{w}_{b,k} + \sigma_u^2 \|\mathbf{w}_{b,k}\|^2},
\label{eq:slnr}
\end{equation}
where the leakage covariance
\begin{equation}
\mathbf{Q}_b(t) \;=\; \sum_{l \neq b} \;\sum_{\substack{j=1 \\ s_{l,j}(t)=1}}^{K_l} \mathbf{h}_{b,l,j}\, \mathbf{h}_{b,l,j}^{H}
\label{eq:leakage_cov}
\end{equation}
sums only over UEs that are \emph{currently active} in neighboring cells. The SLNR-maximizing precoder is the dominant generalized eigenvector of the matrix pair $(\mathbf{h}_{b,b,k}\mathbf{h}_{b,b,k}^{H},\, \mathbf{Q}_b(t) + \sigma_u^2 \mathbf{I})$, and in practice we use a rank-$L$ approximation of $\mathbf{Q}_b(t)$ to keep complexity within the per-TTI budget of an edge node.

The dependence on the \emph{current} scheduling state $\mathbf{s}_{l}(t)$ in (\ref{eq:leakage_cov}) is the central issue. When the BS has only the stale observation $\mathbf{s}_{l}(t-\tau)$, it can either (i) plug the stale observation straight into (\ref{eq:leakage_cov}), which builds the wrong leakage covariance and steers nulls toward UEs that are no longer active while ignoring UEs that became active in the meantime, or (ii) infer the current scheduling state from the delayed history. We pursue the second option.

\subsection{Prediction-Assisted Formulation}

Let
\begin{equation}
\mathcal{H}_{b'}(t) \;=\; \big[\mathbf{s}_{b'}(t-\tau),\, \mathbf{s}_{b'}(t-\tau-1),\, \dots,\, \mathbf{s}_{b'}(t-\tau-T_{\mathrm{win}}+1)\big]
\;\in\; \{0,1\}^{T_{\mathrm{win}} \times K_{b'}}
\label{eq:history}
\end{equation}
denote the most recent $T_{\mathrm{win}}$ delayed scheduling vectors of BS $b'$ as observed at a neighboring BS. Each BS applies a learned per-cell predictor $f_{\theta_{b'}}$ to produce a current-time estimate
\begin{equation}
\hat{\mathbf{s}}_{b'}(t) \;=\; \mathbb{1}\!\left[\, f_{\theta_{b'}}\!\big(\mathcal{H}_{b'}(t)\big) \,>\, 0.5\, \right] \;\in\; \{0,1\}^{K_{b'}},
\label{eq:predictor}
\end{equation}
which replaces $\mathbf{s}_{b'}(t-\tau)$ in (\ref{eq:leakage_cov}). The overall design problem then becomes a joint learning-and-control task: learn per-BS predictors $\{f_{\theta_{b'}}\}$ that, when wired into the SLNR precoder of (\ref{eq:slnr})--(\ref{eq:leakage_cov}), recover as much as possible of the sum rate in (\ref{eq:cbf_obj}) at delay $\tau$.

Two practical constraints follow from the edge setting. First, the predictor must run locally on the BS at TTI granularity, which rules out any centralized or high-latency solution. Second, the predictor should not depend on a fixed UE identity or fixed ordering, since users enter and leave the cell continuously. We therefore require $f_{\theta}$ to be permutation-invariant in the UE dimension, so that the same trained model handles a varying $K_{b'}$ without retraining.

\section{Proposed Spatio-Temporal Scheduling Prediction Framework}
\label{sec:prediction}


The proposed framework is a two-stage pipeline (Fig.~\ref{fig:solution}). In Stage~1, each BS trains a spectral-temporal GNN predictor on its own observed history of (delayed) neighbor scheduling. In Stage~2, the trained predictor is wired into the SLNR coordinated beamformer so that the leakage covariance (\ref{eq:leakage_cov}) is built from $\hat{\mathbf{s}}_{b'}(t)$ instead of $\mathbf{s}_{b'}(t-\tau)$. Both stages share the same simulator, channel realizations, and proportional-fair scheduler, so the predictor is trained on the same operating regime in which it is later deployed.

\subsection{Spectral-Temporal GNN Predictor}
\label{subsec:stemgnn}

We use a StemGNN architecture~\cite{cao}, a spectral-temporal graph neural network originally proposed for multivariate time-series forecasting. The fit is direct: the $K_{b'}$ scheduling sequences of cell $b'$ are correlated through the proportional-fair scheduler that produced them, and StemGNN jointly models temporal and inter-series dependencies in the spectral domain. The network has three components.

\noindent\textbf{Latent correlation layer.} The input window $\mathbf{X} \in \{0,1\}^{T_{\mathrm{win}} \times K_{b'}}$ is passed through a GRU~\cite{cho2014gru}, and a self-attention layer over the hidden states produces a learned adjacency matrix $\mathbf{A} \in \mathbb{R}^{K_{b'} \times K_{b'}}$. This $\mathbf{A}$ captures which pairs of UEs tend to be co-scheduled and is learned from data rather than specified a priori.

\noindent\textbf{StemGNN block.} A Graph Fourier Transform (GFT) using the eigenbasis of $\mathbf{A}$ projects the multivariate input into a spectral domain where inter-UE correlations become orthogonal. A Spectral Sequential Cell then applies a Discrete Fourier Transform along the temporal axis of each (now decorrelated) component, processes the frequency representation with a 1D convolution and a Gated Linear Unit, and inverts back to the time domain. A learned graph convolution and an inverse GFT close the block. Two such blocks are stacked with a residual connection, where the second block learns the reconstruction residual of the first.

\noindent\textbf{Output layer.} A Gated Linear Unit followed by fully connected layers produces per-UE activation logits, which are passed through a sigmoid $\sigma(\cdot)$ and thresholded at $0.5$ as in (\ref{eq:predictor}) to yield the binary scheduling prediction. Since scheduling prediction is a binary classification task rather than the regression task in~\cite{cao}, we train the network with the binary cross-entropy loss
\begin{equation}
\mathcal{L}(\theta) \;=\; -\frac{1}{|\mathcal{D}|} \sum_{(\mathbf{X}, \mathbf{y}) \in \mathcal{D}} \sum_{k=1}^{K_{b'}} \Big[ y_k \log \hat{y}_k + (1 - y_k) \log(1 - \hat{y}_k) \Big],
\label{eq:bce}
\end{equation}
where $\hat{y}_k = \sigma(f_\theta(\mathbf{X}))_k$ and $\mathcal{D}$ is the training set of (history window, target scheduling) pairs.

\subsection{Permutation Invariance and Variable User Population}
\label{subsec:perm_inv}

To meet the second edge constraint above, the input UEs are treated as nodes of the latent graph rather than as fixed positions in a vector. The latent correlation layer infers $\mathbf{A}$ from data on each forward pass, the spectral graph convolution is permutation-equivariant by construction, and the per-UE output head is shared across UE indices. As a result, the same trained model handles a varying number of UEs in cell $b'$ without retraining: UEs that enter the cell are added as new nodes, and UEs that leave are dropped, with no change to the parameters $\theta$.

\subsection{Integration with the SLNR Beamformer}
\label{subsec:integration}

At inference time, each receiving BS $b$ runs $B - 1$ copies of the trained predictor, one per neighboring cell, on the delayed history $\mathcal{H}_{b'}(t)$ it has on hand. The predicted neighbor scheduling vectors $\hat{\mathbf{s}}_{b'}(t)$ are fed into (\ref{eq:leakage_cov}) to build the estimated leakage covariance $\hat{\mathbf{Q}}_b(t)$, and the SLNR precoder of (\ref{eq:slnr}) is then computed via a rank-$L$ generalized eigenvalue decomposition. No other change is made to the beamforming algorithm, which keeps the predictor and the beamformer cleanly decoupled and lets the predictor be retrained or swapped without touching the physical-layer code path.

In operational terms, each BS keeps producing a beamforming decision at every TTI even if neighbor reports arrive late, and the decision is no longer a direct function of the most recent received message. The predictor acts as a learned model of expected neighbor behaviour, so a single missing, late, or anomalous report cannot, on its own, drive the beamformer into a faulty state.

\subsection{Algorithm}
\label{subsec:algorithm}

Algorithm~\ref{alg:pipeline} summarizes the complete pipeline. Lines~3--5 cover the initial data-gathering phase, during which the BSs run conventional CBF-SLNR with stale scheduling and accumulate a history tensor $\mathcal{H} \in \{0,1\}^{B \times K \times T_{\mathrm{total}}}$. At a fixed TTI $t_{\mathrm{train}}$, each BS trains its own predictor on its accumulated history (lines~16--19). From $t_{\mathrm{train}}+1$ onward, the trained predictor replaces the stale neighbor scheduling in the SLNR precoder (lines~7--10). Training uses the Adam optimizer with learning rate $10^{-3}$, mini-batch size $32$, gradient clipping at maximum norm $1.0$, early stopping on validation loss with patience $15$, and a strictly chronological $80/20$ train--validation split that avoids any temporal leakage from future to past.

\begin{algorithm}[!htbp]
\DontPrintSemicolon
\SetAlgoLined
\SetKwInOut{Input}{Input}
\SetKwInOut{Output}{Output}
\SetKw{KwTo}{to}
\Input{Channels $\mathbf{H}$; backhaul delay $\tau$; history window $T_{\mathrm{win}}$; horizon $T_{\mathrm{total}}$; training trigger $t_{\mathrm{train}}$.}
\Output{Beamformers $\{\mathbf{w}_{b,k}(t)\}$; sum rate; edge rate.}
\BlankLine
Initialize history $\mathcal{H} \in \{0,1\}^{B \times K \times T_{\mathrm{total}}}$, PF rate histories $\bar{r}_b(t)$, predictors $\leftarrow \emptyset$\;
\For{$t \leftarrow 0$ \KwTo $T_{\mathrm{total}} - 1$}{
  \tcp{Scheduling}
  \lFor{each BS $b$}{$\mathbf{s}_b(t) \leftarrow \textsc{PF}(\mathbf{H}, \bar{r}_b(t))$; store in $\mathcal{H}[\cdot,\cdot,t]$}
  \tcp{Neighbor scheduling estimate}
  \eIf{$\mathrm{predictors} = \emptyset$}{
    $\hat{\mathbf{s}}_{b'}(t) \leftarrow \mathbf{s}_{b'}(t - \tau)$ for all $b'$ \tcp*[r]{stale baseline}
  }{
    \For{each BS $b'$}{
      $\mathcal{H}_{b'}(t) \leftarrow \mathcal{H}[b',\,\cdot,\, t-\tau-T_{\mathrm{win}}+1\,:\,t-\tau]$\;
      $\hat{\mathbf{s}}_{b'}(t) \leftarrow \mathbb{1}\!\big[ f_{\theta_{b'}}(\mathcal{H}_{b'}(t)) > 0.5 \big]$\;
    }
  }
  \tcp{CBF-SLNR}
  Build $\hat{\mathbf{Q}}_b(t)$ from $\{\hat{\mathbf{s}}_{b'}(t)\}_{b' \neq b}$ via~(\ref{eq:leakage_cov})\;
  Compute SLNR precoder $\mathbf{w}_{b,k}(t)$ via rank-$L$ GEVD of $(\mathbf{h}_{b,b,k}\mathbf{h}_{b,b,k}^{H},\,\hat{\mathbf{Q}}_b(t) + \sigma_u^2 \mathbf{I})$\;
  Compute rates $R_{b,k}(t)$ via~(\ref{eq:sinr_rate}); update $\bar{r}_b(t)$\;
  \tcp{Train predictors once}
  \If{$t = t_{\mathrm{train}}$}{
    \For{each BS $b'$}{
      Build $\mathcal{D}_{b'}$ from $\mathcal{H}[b',\,\cdot,\,0:t_{\mathrm{train}}]$ with lag $\tau$, window $T_{\mathrm{win}}$; chronological $80/20$ split\;
      Train $f_{\theta_{b'}}$ by minimizing~(\ref{eq:bce}) with Adam, BCE loss, gradient clipping, early stopping\;
    }
    $\mathrm{predictors} \leftarrow \{ f_{\theta_{b'}} \}_{b'=1}^{B}$\;
  }
}
\Return $\{\mathbf{w}_{b,k}(t)\}$, sum rate, edge rate\;
\caption{Two-stage prediction-assisted CBF-SLNR pipeline.}
\label{alg:pipeline}
\end{algorithm}

\section{Experimental Results}
\label{sec:results}

\subsection{Experimental Setup}

We evaluate the proposed framework on a three-cell Massive MIMO downlink network with 60 single-antenna UEs (22, 20, and 18 in cells 0, 1, and 2 respectively), 64 antennas per base station, and an identical proportional-fair scheduler running at every BS. Channel matrices are generated with the Quadriga Urban Micro (UMi) propagation model~\cite{dahlman}, which captures path loss, shadowing, and multipath fading under realistic urban conditions and is widely used in 5G and 6G system-level evaluation. The configuration follows parameters representative of operational Massive MIMO deployments rather than a stylised academic setup. Two subcarrier configurations are evaluated, 1~SC (narrowband) and 48~SC (wideband), which together cover the operating regimes typical of real deployments. Backhaul delays of $\tau \in \{1, 3, 5\}$ TTIs are considered. All models use a 12-step delayed history window ($T_{\mathrm{win}} = 12$).

StemGNN is benchmarked against LSTM~\cite{hochreiter1997lstm}, GRU~\cite{cho2014gru}, Simple RNN~\cite{elman1990rnn}, third-order Markov chains, and a Simple Moving Average baseline. Two experiments are conducted. \textbf{Experiment~1} evaluates standalone scheduling prediction over 2000 TTIs with a 70/20/10 chronological train/validation/test split. \textbf{Experiment~2} integrates the trained predictors into the live CBF-SLNR pipeline over 2000 TTIs, with model training triggered at TTI~1600 on the scheduling history accumulated under coordinated beamforming conditions, and performance measured over the final 400 TTIs. The relative improvement of StemGNN over a baseline is reported as $(V_{\mathrm{StemGNN}} - V_{\mathrm{baseline}}) / V_{\mathrm{baseline}} \times 100\%$. Results are aggregated across 18 independent operating conditions, namely three cells with different UE populations, two subcarrier configurations, and three backhaul lags, so the trends below reflect consistent behaviour rather than a single favourable run.

\subsection{Scheduling Prediction Accuracy}

Table~\ref{tab:pred_h1} reports prediction accuracy at Horizon~1 along with StemGNN's relative improvement over each baseline.

\begin{table}[htbp]
\centering
\caption{Scheduling prediction accuracy (\%) at Horizon~1, and StemGNN's relative improvement over each baseline.}
\label{tab:pred_h1}
\begin{tabular}{@{}lccccc@{}}
\toprule
\textbf{Model} & \textbf{BS~0} & \textbf{BS~1} & \textbf{BS~2} & \textbf{Mean} & \textbf{Improv.\ (\%)} \\
\midrule
Simple Moving Average  & 66.59 & 59.36 & 60.47 & 62.14 & $+$40.92 \\
3rd Order Markov       & 86.00 & 83.53 & 86.28 & 85.27 & $+$2.70  \\
Simple RNN             & 86.19 & 85.05 & 86.21 & 85.82 & $+$2.04  \\
GRU                    & 86.34 & 85.34 & 86.33 & 86.00 & $+$1.83  \\
LSTM                   & 86.24 & 84.87 & 86.87 & 85.99 & $+$1.84  \\
\midrule
\textbf{StemGNN (proposed)} & \textbf{89.71} & \textbf{85.71} & \textbf{87.30} & \textbf{87.57} & --- \\
\bottomrule
\end{tabular}
\end{table}

StemGNN reaches 87.57\% mean accuracy at Horizon~1 and leads on every cell. It is ahead of the strongest recurrent baseline (GRU) by 1.83 percentage points, ahead of the third-order Markov chain by 2.70, and ahead of the Simple Moving Average by a much larger 40.92. The recurrent baselines cluster tightly between 85.82\% and 86.00\%, which suggests that gating mechanisms add little over plain recurrence for short-range binary scheduling dynamics. At Horizon~1, the lead of the spectral-graph predictor over a well-tuned recurrent baseline is modest but consistent across cells.

Table~\ref{tab:pred_horizons} reports accuracy at Horizons~3 and~5.

\begin{table}[htbp]
\centering
\caption{Scheduling prediction accuracy (\%) at Horizons~3 and~5, and StemGNN's mean improvement over each recurrent baseline.}
\label{tab:pred_horizons}
\begin{tabular}{@{}lcccccccc@{}}
\toprule
& \multicolumn{4}{c}{\textbf{Horizon~3}} & \multicolumn{4}{c}{\textbf{Horizon~5}} \\
\cmidrule(lr){2-5} \cmidrule(lr){6-9}
\textbf{Model} & BS~0 & BS~1 & BS~2 & \textbf{Improv.}
               & BS~0 & BS~1 & BS~2 & \textbf{Improv.} \\
\midrule
Simple RNN  & 77.52 & 75.29 & 77.54 & $+$6.38\%
            & 76.54 & 72.68 & 76.58 & $+$5.40\% \\
GRU         & 77.44 & 75.04 & 78.33 & $+$5.72\%
            & 75.43 & 72.57 & 76.48 & $+$5.41\% \\
LSTM        & 76.39 & 73.47 & 77.41 & $+$7.71\%
            & 74.39 & 70.91 & 75.88 & $+$7.44\% \\
\midrule
\textbf{StemGNN} & \textbf{83.00} & \textbf{76.00} & \textbf{82.00} & ---
                 & \textbf{80.00} & \textbf{75.00} & \textbf{79.00} & --- \\
\bottomrule
\end{tabular}
\end{table}

StemGNN keeps a 3 to 6 percentage-point lead over every recurrent baseline at both longer horizons, and the margin grows with horizon. The advantage over LSTM widens from 1.84\% at Horizon~1 to 7.71\% at Horizon~3, and the advantage over GRU widens from 1.83\% to 5.72\% over the same range. As short-range temporal autocorrelation decays at longer horizons, the dominant predictive signal shifts to inter-UE co-scheduling structure, and the spectral graph component of StemGNN captures that structure jointly across all UEs while the recurrent models still process each sequence in isolation.

\subsection{CBF-SLNR Integration Results}

Table~\ref{tab:latency_baseline} reports CBF-SLNR performance under stale scheduling with no ML compensation. This setting isolates what happens when the backhaul fault is left uncorrected.

\begin{table}[htbp]
\centering
\caption{CBF-SLNR performance under backhaul latency without ML prediction. Sum rate and edge rate are in bits/s/Hz.}
\label{tab:latency_baseline}
\begin{tabular}{@{}lcccc@{}}
\toprule
& \multicolumn{2}{c}{\textbf{48~SC}} & \multicolumn{2}{c}{\textbf{1~SC}} \\
\cmidrule(lr){2-3} \cmidrule(lr){4-5}
\textbf{Configuration} & Sum Rate & Edge Rate & Sum Rate & Edge Rate \\
\midrule
REZF (uncoordinated)   & 149.40 & 1.13 & 198.61 & 1.37 \\
CBF-SLNR Lag~0 (ideal) & 156.99 & 1.30 & 257.11 & 1.87 \\
CBF-SLNR Lag~1         & 138.93 & 1.12 & 205.89 & 1.51 \\
CBF-SLNR Lag~3         & 146.17 & 1.19 & 222.80 & 1.64 \\
CBF-SLNR Lag~5         & 145.40 & 1.17 & 223.61 & 1.37 \\
\bottomrule
\end{tabular}
\end{table}

One TTI of backhaul delay reduces the 48-SC sum rate by 11.5\% and actually pushes performance \emph{below} the uncoordinated REZF baseline (138.93 vs 149.40), which means that running coordinated beamforming on stale inputs is worse than not coordinating at all. At Lag~1, the precoder nulls towards UEs that are no longer scheduled and steers interference towards UEs that are actually active. Lag~3 partially recovers because information several slots old loses its specificity and the precoder degrades towards generic, uncoordinated beamforming, which is harmless by comparison. This is the scenario in which a deployed system needs a fault-tolerance mechanism that activates when fresh information is missing.

\begin{figure}[htbp]
\centering
\includegraphics[width=0.49\textwidth]{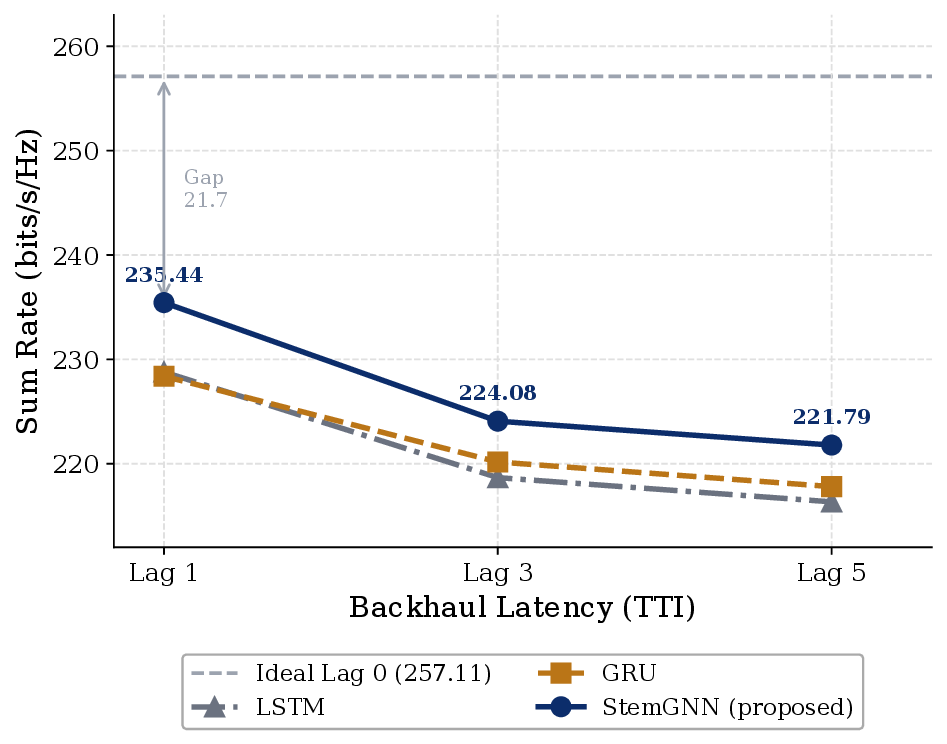}
\hfill
\includegraphics[width=0.49\textwidth]{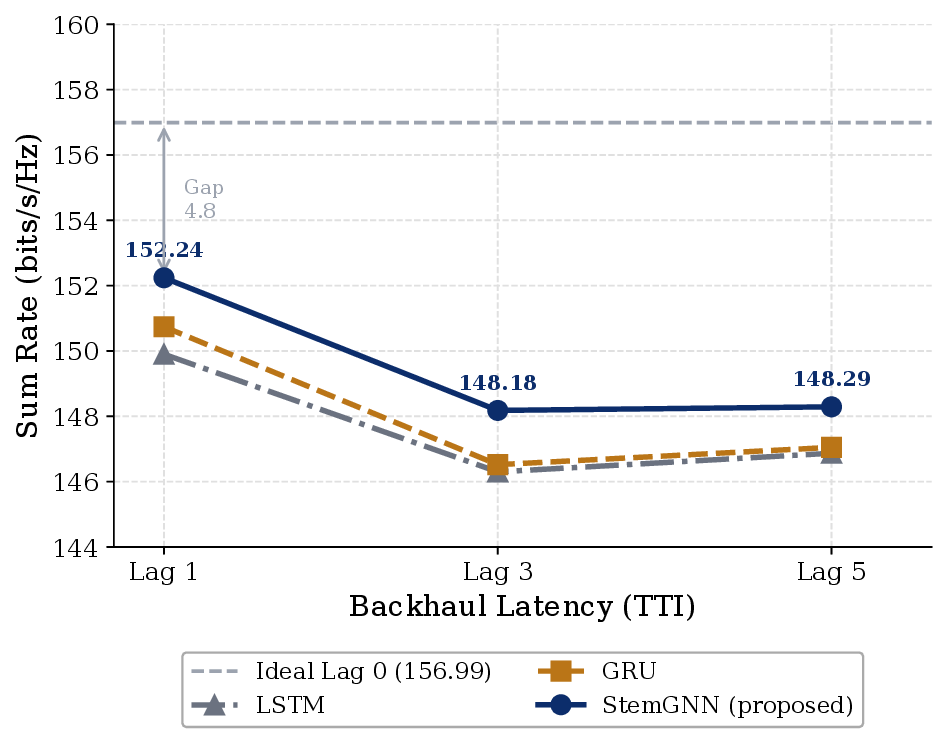}
\caption{CBF-SLNR sum rate versus backhaul latency with ML-predicted schedules for 1~SC (left) and 48~SC (right). StemGNN consistently leads at all lag values. The dashed line marks the ideal Lag~0 upper bound.}
\label{fig:sumrate}
\end{figure}

Table~\ref{tab:cbf_ml} and Fig.~\ref{fig:sumrate} report end-to-end CBF-SLNR performance with ML-predicted schedules at Lag~1, the operating point where the gain is largest.

\begin{table}[htbp]
\centering
\caption{CBF-SLNR performance with ML-predicted schedules at Lag~1. Sum rate and edge rate are in bits/s/Hz, and \emph{Improv.} is StemGNN's relative gain in sum rate over the baseline.}
\label{tab:cbf_ml}
\begin{tabular}{@{}lcccccc@{}}
\toprule
& \multicolumn{3}{c}{\textbf{48~SC}} & \multicolumn{3}{c}{\textbf{1~SC}} \\
\cmidrule(lr){2-4} \cmidrule(lr){5-7}
\textbf{Model} & Sum Rate & Edge Rate & \textbf{Improv.}
               & Sum Rate & Edge Rate & \textbf{Improv.} \\
\midrule
No prediction  & 138.93 & 1.117 & $+$9.58\%  & 205.89 & 1.507 & $+$14.35\% \\
LSTM           & 149.90 & 1.195 & $+$1.56\%  & 228.79 & 1.729 & $+$2.91\%  \\
GRU            & 150.74 & 1.202 & $+$0.99\%  & 228.39 & 1.723 & $+$3.09\%  \\
\midrule
\textbf{StemGNN} & \textbf{152.24} & \textbf{1.283} & --- & \textbf{235.44} & \textbf{1.757} & --- \\
Ideal (Lag~0)  & 156.99 & 1.300 & --- & 257.11 & 1.871 & --- \\
\bottomrule
\end{tabular}
\end{table}

With the StemGNN predictor in the loop at Lag~1, the 48-SC sum rate climbs from 138.93 to 152.24 bits/s/Hz, recovering 73\% of the gap to the ideal Lag-0 upper bound and improving on the no-prediction baseline by 9.58\%. In the 1-SC setting, StemGNN recovers 57\% of the Lag-1 loss with a 14.35\% sum-rate improvement over the no-prediction baseline. The largest gains land at the Lag-1 operating point, which is the same point at which the uncompensated system is most unsafe to deploy. StemGNN also stays ahead of the recurrent baselines under integration: it beats GRU by 0.99\% (48~SC) and 3.09\% (1~SC), and LSTM by 1.56\% (48~SC) and 2.91\% (1~SC). The advantage narrows at higher lags as all models converge towards similar accuracy plateaus, which Fig.~\ref{fig:sumrate} shows clearly.

The edge-rate gains are larger in relative terms than the sum-rate gains, and we consider this one of the more important findings for an edge-availability setting. In the 1-SC configuration, StemGNN raises the edge rate from 1.507 to 1.757 bits/s/Hz at Lag~1, a 16.59\% improvement, and recovers 83\% of the Lag-1 fairness loss. The improvements persist at Lag~3 and Lag~5 even as the sum-rate benefits start to fade, which says that the framework keeps service quality consistent for the cell-edge users who are most exposed to coordination failure~\cite{tse2005wireless}. In an edge deployment, this translates directly into more even quality of experience across cells when the backhaul is degraded.

\begin{figure}[htbp]
\centering
\includegraphics[width=0.65\textwidth]{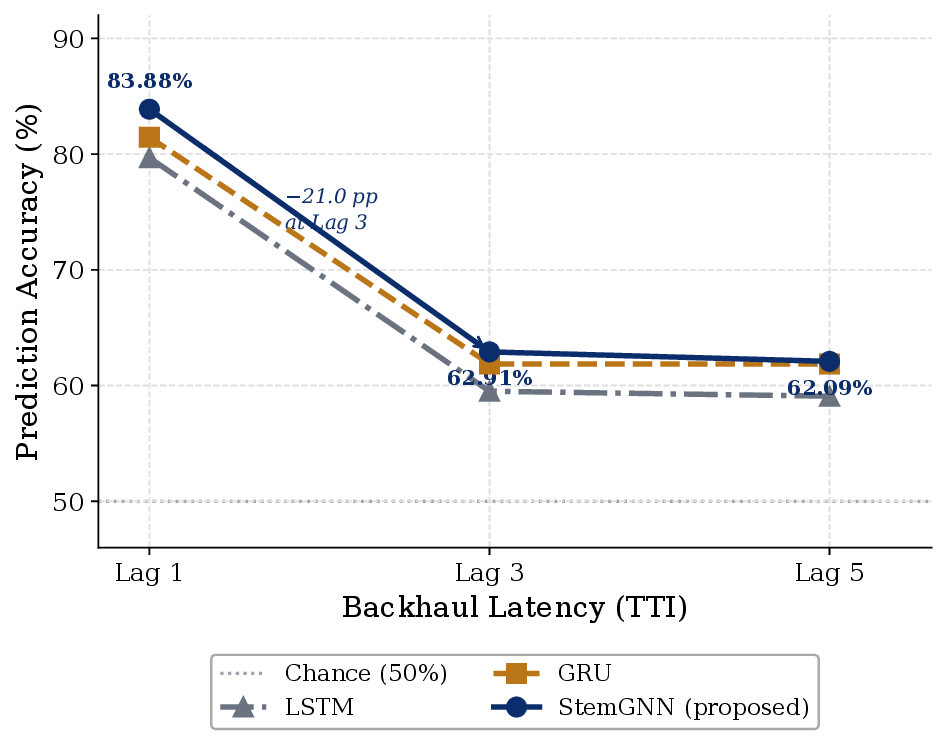}
\caption{Scheduling prediction accuracy during CBF-SLNR integration (48 subcarriers). All models drop sharply from Lag~1 to Lag~3 and plateau thereafter. StemGNN leads at Lag~1 with 83.88\%.}
\label{fig:accuracy}
\end{figure}

Fig.~\ref{fig:accuracy} reports prediction accuracy during live CBF-SLNR integration across backhaul latency values. All models drop steeply from Lag~1 to Lag~3, a drop of 21 percentage points for StemGNN (from 83.88\% to 62.91\%), and plateau thereafter. At Lag~1, StemGNN leads GRU by 2.42 pp and LSTM by 4.19 pp. The collapse in accuracy at Lag~3 directly explains the shrinking sum-rate benefit visible in Fig.~\ref{fig:sumrate}. The integrated accuracy at Lag~1 (83.88\%) is slightly below the standalone accuracy at Horizon~1 (87.57\%). This gap reflects a closed-loop distributional shift: the predictor's own influence on the beamformer alters the interference environment and subtly shifts the scheduling dynamics away from the patterns it was trained on. This is a known property of any closed-loop learned controller rather than a defect of the model itself.

\subsection{Discussion}

Three findings that stand out are as follows:

\textbf{First}, the latency-performance relationship is non-monotonic. Lag~1 is worse than Lag~3 or Lag~5 because the precoder nulls towards users active in the previous slot who may no longer be active, so confident misinformation turns out to be worse than no information at all. The proposed framework directly addresses this failure mode and recovers 57 to 73\% of the Lag-1 sum-rate loss with 9.58 to 14.35\% improvement over the no-prediction baseline. From a fault-tolerance perspective, this is the worst-case window the edge controller must survive, and it is the window in which the predictor delivers its largest gain.

\textbf{Second}, StemGNN's margin over recurrent baselines grows with prediction horizon, reaching up to 7.71\% over LSTM at Horizon~3. This confirms that inter-UE co-scheduling structure becomes the dominant predictive signal once short-range autocorrelation decays. The numbers reported here should be read as conservative estimates. The Quadriga UMi simulator with a proportional-fair scheduler produces scheduling patterns that are more regular than those of a real network with bursty traffic, high user mobility, denser deployments, and many more UEs per cell. In those more realistic conditions, the structural inter-UE dependencies that the spectral graph component captures would carry substantially more predictive value, and we expect the gap over purely temporal recurrent models to widen.

\textbf{Third}, accuracy is high at Lag~1 (83.88\% for StemGNN) but drops by roughly 21 percentage points at Lag~3 and plateaus thereafter. Two effects compound here: the growing uncertainty of multi-step forecasting, and the closed-loop distributional shift noted above. This accuracy collapse is the main bottleneck of the current framework and is what limits sum-rate recovery at Lag~3 and Lag~5. The shape of this failure profile is itself useful information for an edge deployment: the system is most reliable exactly at the operating point where it most needs to be reliable, and degrades gracefully towards the uncoordinated baseline as latency grows further, rather than collapsing into a faulty state. Across all 18 operating conditions evaluated, the ranking of StemGNN over the recurrent and statistical baselines remains consistent, indicating that the gains reflect a structural property of the framework rather than an artefact of a specific configuration.

\section{Conclusions}
\label{sec:conclusion}

This paper addressed the problem of coordinated beamforming under backhaul latency in distributed 5G networks, framed as a fault-tolerance problem for edge AI controllers that must operate on stale information from their peers. We proposed a two-stage predictive framework in which a spectral-temporal GNN predicts future UE scheduling states from delayed historical observations, and the predictions replace stale inputs to the CBF-SLNR precoder without modifying the core beamforming algorithm.

Three conclusions stand out. First, the latency-performance relationship is non-monotonic. One TTI of delay is more harmful than three or five, because the precoder actively suppresses interference toward users that are no longer active. Confident misinformation is worse than no information. Prediction is most valuable at precisely this hardest operating point, where the proposed framework recovers 57 to 73\% of the Lag-1 sum-rate loss and brings performance close to the ideal upper bound that assumes zero delay. The framework therefore restores service availability exactly when the system is most exposed to backhaul faults.

Second, spatio-temporal graph modelling captures inter-UE co-scheduling dependencies that purely temporal recurrent models cannot. The advantage grows with prediction horizon, which confirms that structural inter-UE correlations become the dominant predictive signal once short-range autocorrelation decays. The gains reported here are conservative, and in larger and more bursty deployments with greater user mobility, the structural advantage of the spectral graph component is expected to widen.

Third, the performance benefits fall disproportionately on cell-edge users. The framework recovers 83\% of the Lag-1 fairness loss in the 1-SC configuration, which makes it a fairness mechanism for the most exposed users as much as a throughput-recovery tool. For an edge deployment, this means more even quality of experience across cells under the same backhaul-fault conditions.

Future work will target the sharp accuracy drop from Lag~1 to Lag~3 through uncertainty-aware prediction, in which the model outputs a confidence estimate that the precoder uses to weight neighbour scheduling states by reliability. This also extends the trust argument from the Introduction by shrinking the surface exposed to late or anomalous reports. Further directions include end-to-end joint training under a task-aware loss, online adaptive learning to handle closed-loop distributional shift, and evaluation on larger network topologies.

\begin{credits}

\subsubsection{\ackname}
The authors thank Huawei Sweden R\&D for providing the simulation
environment and industrial supervision. This work was carried out
within the Master's thesis programme at the Department of Computer
and Systems Sciences, Stockholm University.

\subsubsection{Use of LLMs.}
Claude (Anthropic) was used for language assistance and drafting under author direction; all technical content is the authors' original work.

\subsubsection{\discintname}
This work was conducted in collaboration with Huawei Sweden R\&D. The beamforming and scheduling simulation framework contains proprietary components and was used solely for academic research purposes. Authors A.~H.~G\"{o}kceoglu and L.~Wang are employed by Huawei. The authors have no other competing interests to declare.

\end{credits}
%
%
%

\bibliographystyle{splncs04}
\bibliography{biblio}

\end{document}